# Analysing the cultural dimensions of cybercriminal groups - A case study on the Conti ransomware group

**Konstantinos Mersinas[1], Aimee Liu[1], Niki Panteli[2]**

[1] *Information Security Group, Royal Holloway, University of London, Egham, Surrey, TW20 0EX, UK*

[2] *Department of Management Science, Lancaster University, Bailrigg, Lancaster, LA1 4YX, UK*

**Abstract**

Cybercriminal profiling and cyber-attack attribution have been elusive goals world-wide, due to their effects on societal and geopolitical balance and stability. Attributing actions to a group or state is a complex endeavour, with traditional established approaches including cyberthreat intelligence and analysis of technical means such as malware analysis, network forensics, and geopolitical intelligence. In this paper, we propose an additional component for profiling cybercriminal groups through analysing cultural aspects of human behaviours and interactions. We utilise a set of variables which determine characteristics of national and organisational culture to create a cultural 'footprint' of cybercriminal groups. As a case study, we conduct thematic analysis across the six dimensions of the Hofstede national culture classification and the eight dimensions of the Meyer classification on leaked internal communications of the ransomware group Conti, a threat for Europe and globally. We propose that a systematic analysis of similar communications can serve as a practical tool for a) understanding the modus operandi of cybercrime and cyberwarfare-related groups, and b) profiling cybercriminal groups and/or nation-state actors. Insights from such applications can, first, assist in combating cybercrime and, second, if combined with additional cyber threat intelligence, can provide a level of confidence in nuanced cyber-attack attribution processes.

1. **Introduction**

The idea of incorporating national culture in cybersecurity is relatively new. Previous research has been conducted in this direction, e.g., in combination to human aspects of risk assessment frameworks (Henshel et al., 2016). Existing research has drawn frome the theoretical and experimental works of Nisbett (2004), Heinrich (Heinrich et al., 2010), and Hofstede (2011). In our study we additionally draw on the work of Meyer (2015). Assuming a simplified attacker-defender dichotomy for the economy of the arguments, the main position of the paper is that the study and analysis of cultural dimensions of cybercriminal groups can be valuable for defenders. The reason is that such analysis can assist in predicting attacking styles, the level of persistence, possible intrusion/attack vectors, and reactions to dynamic defensive responses (Henshel et al., 2016).

The goal of this research is to advance the idea of embedding culture in cyber threat intelligence (CTI), cybercriminal profiling, and cyber-attack attribution. We seek to provide a proof of concept on how analysis of national and organisational culture within cybercriminal groups can be methodologically performed. The analysis is intended to be part of a bigger 'apparatus' which will encompass and combine, along with culture, personality, technical and other CTI information to form broader insights on the modus operandi, constitution, and attacking style of such cybercriminal groups.

The possibility of identifying who the actors behind a cyber-attack are, can serve as a deterrence measure for cybercrime. Such information can be useful for law enforcement and governments for the purpose of protecting critical infrastructure, businesses, the economy, citizens and society more broadly.

The rest of the paper is structured in the following fashion. In section 2 we present the underlying notions utilised in the paper, namely, security culture, the diversification of thinking patterns across cultures, the cultural dimensions of Hofstede and Meyer, criminal profiling, and attribution. Linking these notions provides useful insights for the bigger picture and the potential of this research. In section 3 we examine the evolution and practices of the Conti ransomware group, as the analysed datasets constitute this group's internal communications. Section 4 describes the methodology followed and section 5 presents the analysis and findings. We discuss the findings and hypothesise on their possible explanations in section 6, and conclude in the final section.

## 2. Background

**Security culture VS culture**

Security culture can be defined as the set of human behaviours in an organisational context with the objective to protect information (Da Veiga et al., 2020). It needs to be examined in the context of observable behaviours and it encompasses policies, communication, norms, beliefs, awareness, training, and education. At a higher level, within which security culture exists, we have national and organisational culture. In this paper, we do not examine security culture per se, but these higher levels of culture, which are linked to security culture, since components such as communication, beliefs, and norms within organisational entities (including cybercriminal ransomware groups) are influenced by the overarching culture.

**Cultural dimensions**

We draw on Hofstede's and Meyer's cultural classifications with six and eight cultural dimensions respectively (Minkov and Hofstede, 2011; Meyer, 2015). Effects of these dimensions are observed in individual behaviours and, as an extension, in the environments in which individuals operate. In more detail, the broader notion of 'culture' is observable in these two environments: nations and organisations. Manifestation of the cultural dimensions is not identical in these environments, but there is a correlation. For organisations, in particular, the focus is on how individuals relate to one another, and how the overall organisation functions compared to other organisations. We should note that measuring cultural dimensions is more accurate in organisations, in comparison to nations, as the former have clearer objectives and modes of operation (Hofstede, 2011).

The six dimensions of Hofstede are briefly described first. Applied at societal level, the **Power Distance Index (PDI)** captures the degree to which a society accepts that power is not equally distributed amongst its members. The usual example is the status and the role of politicians or law enforcement within a society. So, individuals from a society with a high score in power distance might maintain this power distance within a group, e.g. with leadership teams being more authoritative towards the low-ranking members.

The **Uncertainty Avoidance Index (UAI)**, then, describes the level of aversion towards uncertainty and ambiguity. It is hypothesised that societies which score low on uncertainty avoidance, tend to be more relaxed, tend to be more practical, and can be less focused on inflexible principles, in comparison with societies which score high on uncertainty avoidance.

Then we have **Individualism versus Collectivism (IvC)**. The first notion, individualism, indicates a view by which individuals focus predominantly on themselves and their immediate family, but not necessarily on the broader community. Then, the other end of this continuum is collectivism, where it is expected, as a norm, that members of a community take care of each other.

The next dimension is **Masculinity versus Femininity (MvF)**. Masculinity characterises societies which are focused more on accomplishments and achievements, and societies which value assertiveness. Being goal-driven and competitive is accepted as a norm in such societies. At the other end, femininity implies a tendency to value cooperation, modesty, and protection of the weakest members. Thus, the norm in such societies is an environment of agreement and collaboration. Again, such attributes can be depicted in the way that organisations and groups operate.

**Long-Term Orientation versus Short-Term Orientation (LvS)** is a dimension which indicates the way a society is connected with its past and history. How valuable traditions are perceived to be, is one manifestation of this dimension and another one is the characteristic of openness to societal change, and by extension, to organisational change. The two ends of this continuum are also referred to as short-term normative, with a suspicion towards change, and long-term pragmatic, with an encouragement of effort as an investment for the future.

Finally, the dimension of **Indulgence versus Restraint (IvR)** attempts to capture, at the one end, an environment which allows gratification, as a measure related to the enjoyment of life. The other end, that is, a higher score on restraint, indicates that the norm within a society is gratification suppression which is usually controlled by social norms.

Although manifestations of these dimensions are quite complex and not straightforward to interpret as phenomena, measurements in dozens of countries around the world indicate different indexes across these dimensions. In a sense, although there are debates about this approach and its validity, one can claim that the set of the relative scores on these dimensions indicates a specific 'footprint' per country. As an example, on average, we observe lower scores of power distance in Europe, in comparison to Asia or Africa; but we identify differences in power distance within Europe, e.g., between south-eastern and north-western Europe. Similarly, these dimensions can be depicted on organisations and on the way they operate. And the point is that cybercriminal groups, especially the larger ones which are involved with cyber organised crime, might be influenced in their modus operandi and structures by these national dimensions, and especially if one or two national cultures are dominant within the group. Or it can be the case that the leadership of a group imposes certain cultural aspects on the other members.

So, these measurements are meant to indicate a relative position amongst countries or societies, across the dimensions and they can potentially provide useful insights, as to how a group operates.

The second set of dimensions, the one suggested by Meyer (2015), has some overlapping variables with Hoftsede's national dimensions to an extent, but they are more oriented towards organisational culture. In this sense, they might be used to describe operations and structures within a group and therefore they are more relevant to the purpose of our study.

The first Meyer dimension is **Communicating (LvH)**. The ends of this continuum are low- and high-context. **Low-context** cultures value communication which is precise, simple, explicit, and clear. In such environments, conveyed messages are understood at face value. Repetition is accepted since it can clarify meaning, and putting messages in writing is also valued. In **high-context** cultures, at the other end, communication tends to be sophisticated, nuanced, and multi-layered.

Then there is the dimension of **Evaluating**, regarding **direct** or **indirect feedback (DvI)**. The measurement here is whether honesty is valued over diplomatic feedback, especially when it comes to negative feedback. Several cultures, for example, prefer to balance negative with positive feedback, or, even consider direct, negative feedback as insulting and unacceptable.

The **Leading** dimension is a measure of the degree of respect towards authority. Therefore, societies or environments can be compared on a continuum from **egalitarian** to **hierarchical (EvH)**. This dimension is similar to Hofstede's variable of power distance, and the cultural studies of leadership by House(House, 2004). In fact, Meyer intended this dimension to be the equivalent of Hofstede's power distance, tailored for businesses.

The dimension of **Deciding (TvC)**, then, measures the degree to which a culture accepts **top-down** decision-making or whether it values **consensus** in reaching decisions. In top-down deciding societies decisions are made at a higher level, e.g., by the leadership team or senior management , and they move "down" to lower-ranking individuals or employees. Both the dimensions of Leading and Deciding are related to the aforementioned variable of power-distance.

The dimension of **Trusting (TvR)**, then, has on the one end a **task-based** culture, where trust is built mostly cognitively through work and accomplishments, and at the other end, we have **relationship-based** environments, where the basis of trust is linked to effective connections and manifests on a more personal level. So, a trustworthy individual within a task-based environment, is one that forms successful collaborations, and prove themselves reliable. In relationship-based cultures, building trust is a process of spending more time and personally getting to know individuals.

The dimension of **Disagreeing (TvI)**, indicates the degree of tolerance for open disagreement. We can have a highly **tolerant** or an **intolerant** environment for open disagreement in an organisation.

The next dimension, **Scheduling (LvF)** describes how much value is placed on operating in a structured, linear fashion, versus being flexible and reactive. Cultures and environments on the two ends of this variable have **Linear**-time scheduling and **Flexible**-time scheduling.

Finally, **Persuading (PvA)**, is a dimension which assesses the ways in which people persuade others and the types of arguments they themselves find convincing. We have the ends of this continuum being called **principles**-first (or deductive) arguments and then **applications**-first (or inductive) arguments. Studies indicate that southern European and Germanic cultures tend to value **deductive** arguments – or **principles-first** arguments - as mostly persuasive. American and British cultures, then, are more likely to utilise **inductive** logic – or **applications-first** logic. Interestingly, this dichotomy is reflected in the legal systems of the aforementioned countries and geographical areas, namely, as common law (inductive, e.g., in the US and UK) and civil, constitution-based, law (deductive, e.g., in Europe and Russia).

**Culture as a predictor of thought**

Nisbett identifies a high level dichotomy between Eastern and Western patterns of thinking. Through experimental research it is shown that these two broad cultures shape individuals' thinking patterns and perceptions of the world. As an indicative example, Eastern thought patterns tend to utilise relationships and contextual parameters, whereas Western thought focuses more on objects' characteristics in isolation from their environment. Research, e.g., indicates that toddlers in the West learn to talk through learning nouns, as opposed to toddlers in the East who mostly learn verbs (Nisbett, 2004).

With regards to cybersecurity, in particular, specific components of attacks, such as deception, have been identified to be influenced by cultural biases (Almeshekah and Spafford, 2014). Moreover, our focus is on cybercriminal groups, where norms, peer observation, and benchmarking is prominent. Additionally, status elevation, and the building of a reputation within the community or group are factors which are influenced by the established culture within the group. The same behaviours can be either celebrated or criticised depending on the established cultural norms of the group (Minkov et al., 2011).

**Profiling**

Cybercriminal profiling approaches have been relatively less studied in comparison to traditional crimes. Even in the few attempts for such an endeavour, the component of culture is missing, or at least is not utilised in the fashion that we propose. Lickiewicz (2011) proposes intelligence, personality, technical skills, internet addiction, and social skills as the main variables (Lickiewicz, 2011), in one of the most thorough hacker profile models, to the best of our knowledge. The model's component which is mostly related to the proposed cultural component is *social skills,* which is, however, depicted as closely-related to personality traits[1], as it encompasses the internalisation of social norms, the adaptation to these norms, and the individual's relationship formation. This is in contrast to the established social and cultural norms themselves, in a cybercrime environment.

---

[1] We consider personality traits in line with the Five Factor Model of personality in the seminal work of Costa and McCrae (1991) and further related studies, such as De Young et al. (2009).

Profiling techniques are categorised into inductive and deductive. Inductive techniques for cybercriminal profiling involve the *'study of a group of subjects who share a common characteristic or activity'* and the goal is to identify characteristics and patterns in their behaviour (Shaw, 2006). The approach can be qualitative or quantitative, the latter include statistical methods on large datasets to reveal factors that allow for patterns to be identified (Shinder et al., 2002). On the other hand, deductive profiling techniques start from the 'crime scene' (with its extended notion in cybersecurity), collect evidence on the victim, and try to identify the motivations of the offender. Thus, deductive approaches refer to an *'assessment of a subject's personal characteristics from his or her crimes, activities, statements or other reports'* (Shaw, 2006); that is, the characteristics of the attacker are deduced (Tennakoon, 2011). Deductive approaches are case-oriented and include the profiler testing a number of hypotheses on the existing information, in line with our cultural proposal. There appears to be a preference for deductive techniques in the literature, since they allow for studies based on specific cases by utilising the characteristics, personality traits, history, and other variables of the offenders.

As an extension to profiling, we add cyber threat intelligence (CTI) and cyber-attack attribution. Pleskonjić et al. (2006) propose the hypothesis that a collection of criminal characteristics can be used to identify the motivations and the culture that the criminal grew up in (Pleskonjić, Đorđević, Maček and Carić, 2006, as cited in Lickiewicz, 2011).

**On Attribution**

We can have a number of attacks claimed to come from specific actors. These claims can be driven by the media, or governments, according to their geo-political agendas. There can also be attempted deception from the actors themselves, in order to blame another state or group, or to cover their trail. In any case, cyber-attack attribution is not an exact science (Rid and Buchanan, 2015), and thus, the idea here is that a collection of diverse evidence can shed light on, or, at least provide useful insights about the true actors behind the attacks.

Thus, our approach aims in showcasing insights from the cultural component of cybercrime, and indicates the direction of linking culture with cybercriminal profiling models for understanding cybercriminal behaviours and providing hints on cyber-attack attribution.

## 3. Case study: the Conti Ransomware Group

Ransomware refers to the operating procedure under which an attacker asks for ransom to solve issues, which they create by encrypting files or making systems unusable/inaccessible in the first place, by using the corresponding malware (Kumar and Ramlie, 2021).

The group which came to be known as the Conti ransomware group, became very successful in the ransomware ecosystem and economy. They are found to have started operations in 2020, and rebranded two years later. Identified malware within the group include Emotet, TrickBot and Ryuk. The typical modus operandi of the group was the request for ransom and double extortion, via data exfiltration and blackmailing of targets to leak this data.

It is reported that the group changed its brand and operations, due to the complications which followed their public support to the Russian government regarding the war in Ukraine (Kovacs, 2022). Cyber Threat Intelligence from private companies indicates that the group's senior leadership might have had contact with Russian agencies (Burgess, 2022), however, the group is found to be working largely on their own. Following the support to Russia, a Ukrainian researcher, who had infiltrated the group (Lyngaas, 2022) leaked some 60K internal communication messages of the group (Kovacs, 2022). This is the dataset that we have used for our analysis.

Many of Conti members self-reported to be based in Russia, but the group is linked to Eastern Europe and beyond, e.g., with members in Belarus, Ukraine, and other countries (Burgess, 2022),

Before exiting the 'market', Conti delivered an attack on Costa Rica, stating their goal as the overthrowing of the government; this action is, however, perceived as a distraction during which the group migrated into smaller groups (BleepingComputer, 2022). The attack on Costa Rica resulted in the disruption of the country's customs and tax platforms.

The group is reported to have switched their structure into a decentralised entity with several independent sub-groups (such as Karakurt, Black Basta and BlackByte), operating autonomously. The business model of these subgroups is said to have changed from ransomware to data exfiltration and extortion. Other subgroups continue to operate with locker malware (AlphV (BlackCat), HIVE, HelloKitty (FiveHands), and AvosLocker). The model is not a typical ransomware-as-a-service, since resources and tools are shared amongst the subgroups or affiliate members. It is also reported that the initial attack vector is spear-phishing, following Open Source Intelligence (OSINT) gathering.

Conti targets include several western and European countries. The US government offered a total of $15M for information related to the location or leadership, and assistance for the arrest of the group. Google's Threat Analysis Group (TAG) reported on a threat actor in September 2022, having similarities with a threat actor identified by the Ukrainian CERT (CERT-UA), and namely, UAC-0098 (Huntley, 2023). TAG's conclusion is that former Conti members joined or formed UAC-0098, the objective of which is shown to be attacking Ukrainian targets.

**A more detailed structure of Conti**

We have several indications about the internal structure of the group. First, all members use nicknames or monikers. The head is identified as an individual called 'stern', then 'mango' is a manager who is very active in the leaked datasets.

Within the group, there are identified roles, such as coders (who develop malware and integrate different technologies), testers (who test the malware against security products), administrators (who set up and maintain servers and infrastructure), reverse engineers (who explore code to identify vulnerabilities), and pen-testers (who conduct the actual hacking operations). Spamming operations are believed to have been outsourced, due to a lack of mention in the datasets (krebsonsecurity, 2022.).

## 4. Methodology

The dataset contains daily activities and operations of the group, such as relationship building, members' complaints, communication of tasks, work delegation, motivation, business model and expansion, recruitment, decision-making approaches, salaries, and hierarchical structures. We conducted thematic analysis on the leaked communications across categories relating to everyday operations and relationships in the group.

The cultural dimensions proposed by Hofstede constitute a quantitative framework for measuring cultural values. However, we do not utilise the quantitative aspect of the proposed indexes. We instead combine Hofstede's six dimensions with the eight dimensions of Meyer and conduct qualitative, thematic analysis on the leaked communications of the Conti ransomware group. The reason for the non-quantitative approach is that these values should not be used as absolute values, but instead as indications of a relative position of cultural indexes, across the continuum of each dimension.

We identify communications across these 14 cultural dimensions, codify and categorise messages according to an adjusted thematic analysis methodology of Erlingsson and Brysiewicz (2017). In more detail, our categorisations are checked against the combinations of dimensions as per **Table 1**. The reason for this approach is two-fold. First, we maintain the valuable information from potentially identified individual dimensions, e.g. high scores of *PDI* are identified, indicating that the group's dominant culture has such a characteristic. Second, we examine whether 'groupings' of dimensions indicate a theme, e.g. *ideology* can be examined as a combination of collectivism, high PDI and UAI, femininity, and restraint.

| | | Ideology | Ethical Attributes | Connect-edness | Risk Attitude | Flexibility |
|---|---|---|---|---|---|---|
| **Hofstede's dimensions** | **Code** | | | | | |
| Power Distance Index | PDI | | ✓ low | | | ✓ low |
| Individualism VS Collectivism | IvC | ✓ C | ✓ C | ✓ C | | ✓ I |
| Masculinity VS Femininity | MvF | ✓ F | | ✓ F | ✓ | ✓ |
| Long-term VS Short-term Orientation | LvS | | ✓ L | ✓ L | ✓ | |
| Indulgence VS Restraint | IvR | ✓ R | | | | |
| Uncertainty Avoidance Index | UAI | ✓ high | | | ✓ | |
| **Meyer's dimensions** | | | | | | |
| Persuading: Principles-first (deductive) VS Applications-first (inductive) | PvA | ✓ P | | | | ✓ A |
| Scheduling: Linear-time scheduling VS Flexible-time scheduling | LvF | | | | ✓ | ✓ F |
| Disagreeing: High tolerance for open disagreement VS Low tolerance | TvI | ✓ I | | ✓ T | | ✓ T |
| Trusting: Task-based VS Relationship-based | TvR | | ✓ R | ✓ R | ✓ | ✓ T |
| Deciding: Top-down VS Consensus-minded | TvC | ✓ T | ✓ C | ✓ C | | ✓ |
| Leading: Egalitarian VS Hierarchical | EvH | | ✓ E | ✓ E | | ✓ E |
| Evaluating: Direct negative feedback VS Indirect negative feedback | DvI | | ✓ I | ✓ D | | |
| Communicating: Low-context VS High-context | LvH | | | ✓ L | | ✓ |

**Table 1**: Hofstede's and Meyer's cultural dimensions (Hofstede, 1980; Hofstede, 1983; Hofstede, 1991; Meyer, 2015).
**Hofstede dimensions.** PDI: Power Distance Index; IvC: Individualism vs Collectivism; MvF: Masculinity vs Femininity; UAI: Uncertainty Avoidance Index; LvS: Long-term vs Short-term orientation; IvR: Indulgence vs Restraint.
**Meyer dimensions.** PvA: Principles-first (deductive) VS Applications-first (inductive) persuasion; LvF: Linear-time scheduling VS Flexible-time scheduling; TvI: High tolerance for open disagreement VS Low tolerance for disagreement; TvR: Task-based VS Relationship-based trust; TvC: Top-down VS Consensus-minded decision-making; EvH: Egalitarian VS Hierarchical leadership; DvI: Direct negative feedback VS Indirect negative feedback (Evaluating); LvH: Low-context VS High-context communication.
Individual letters indicate the dimension value related to the construct in the top row.
✓: The dimension is related to the construct in the top row, but the direction of the relationship can go both ways.

Such indications can reveal objectives, motivations, and the overall stance of a cybercriminal group. In more detail, we utilise the mapping of the Hofstede dimensions on the Human Factors Framework (HFF)

(Henshel et al., 2015; Oltramari et al., 2015), as proposed by Henshel et al. (2016), and we extend this mapping to incorporate Meyer's cultural dimensions.

Namely, the mapping includes the themes of ideology, flexibility, connectedness, risk attitude, and ethical attributes as combinations of cultural dimensions or constructs. We omit the remaining two themes of personality traits and motivation. Personality traits refer to relatively stable phenotypic characteristics of thinking patterns and behaviours of individuals, which can be compared and contrasted. These, however, have both environmental and biological causes, thus, they might not be in accordance with the established culture. Motivation, on the other hand, is influenced by all other themes, so utilising it as a separate grouping would not provide additional value.

In particular, **ideology** relates to a high degree of uncertainty avoidance, being restrained and having self-control, as opposed to seeking personal gratification, it is based on principles which might not allow space for disagreement, and has a top-down nature in decision-making. Ideology reflects the objectives and the motivations of the group. Similarly to flexibility, a pragmatic, application-oriented stance which allows open disagreement, that is a non-ideological orientation is more in line with a business-driven group. Threat actors often portray their motivations in a deceptive fashion, thus, identifying ideological components can clarify a group's true intentions.

**Flexibility** refers to the degree of adaptability and the willingness to change; it is linked with low power distance, individualistic and pragmatic-oriented environments. Flexibility can relate to most of Meyer's dimensions, e.g., applications-first arguments are more adjustable to the circumstances, scheduling, tolerance for disagreement can allow for adaptability, trust that is task-based, an egalitarian and low-context communication can be conditions for flexible behaviour. Flexibility reflects the business model of the group, how they react to obstacles, and how they adapt to the 'market', especially for cybercriminal groups who draw international law enforcement attention.

**Ethical attributes** are related to a sense of what is acceptable or not, under a given belief system or social norms. PDI influences such attributes since dominant individuals can impose their beliefs, but equally, low-power individuals might justify their actions due to their position. The latter case is a manifestation of the 'Robin Hood effect' where cyber-attacks from less developed countries towards western countries, which support the local communities, are justified and rationalised.

**Connectedness** refers to the bonding and the closeness of relationships that individuals portray within a group. It can refer to collaborations and more personal relationships, and indicates a sense of community. Collectivism is directly related to connectedness, since community members take care of each other. Femininity is also directly linked to connectedness, through valuing cooperation and member protection. However, environments portraying masculinity can still allow for a sense of community. Long-term oriented environments, then, allow for a holistic worldview, including the evolution and progress of the group or community.

**Risk attitude** refers to manifested risk-seeking or risk-averse behaviours. Masculinity is a risk-seeking indicator, since antagonistic environments promote risk-taking actions, whereas femininity promotes risk aversion, similarly to high scores in UAI. Long-term orientations also promote patience and less overt actions, including a prolonged deployment of cyber-attack campaigns with smaller probability of exposure.

We utilise the three leaked datasets of the Conti internal communications found at GitHub (TheParmak, 2023). These are categorised into three years, from 2020 to 2022 and they consist of daily communications amongst members, either in pairs or in group discussions.

Our analytical approach was informed by Braun and Clarke (2006) thematic analysis which included familiarisation with the data, initial code generations, theme search and identification. The analysis was conducted by two of the authors, and this initially involved the categorisation of the messages across the

14 dimensions of our theoretical framing, first independently for the purposes of triangulation and then collectively. Following from this, we sought to identify core emerging themes; these were notably **connectedness** and the power dynamics of the group, with an increased **power distance.**

## 5. Findings

In this section we present the subset of the 14 dimensions which is identified in the datasets. We provide indicative quotations to highlight the relevant findings, and in doing so we also provide the timestamp, which indicates the date and time of the communicated message, as a reference to the particular communication within the datasets.

First, we identify indications for a wider sense of collaboration, that is for **collectivism**, which refers to the integration of the individuals in the group. There are several messages that express wishes for holidays and vacations. Then, knowledge sharing is an aspect that conveys the common goals of the group, e.g., the member 'revers' tells another member *'we'll teach you everything in terms of practice and everything else'* (2021-05-11T08:45:44.967012), 'mango' mentions *'tell me what you need from me and we'll get it done'* (2021-08-30T05:21:59.601158), the member 'salamandra' states *'bro, when you get your paycheck and you know it's okay, give me some more coders'* (2020-07-01T18:38:24.304875"), and the member 'tramp' asks *'guys when they pumped out the date, couldn't they have left traces of IP addresses somewhere?'* (2022-01-27T20:34:20.665620). The above indicate an overall collaboration and a sense of 'we' in terms of the group's goals. The group member called 'target' addresses 'stern' (the leader) to say that they did a task 'all by myself, unfortunately'. Maybe this indicates a personal complaint and a self-portraying of achievement to the leader. But, overall, individuals within the group seem to be working towards overarching objectives, similarly to a well-functioning business. For example, the structure of the group - or organisation - is observed to work well, with many group members voluntarily committing to the objectives.

With regards to **leadership** style, i.e. egalitarian or hierarchical, and with regards to **deciding**, i.e. top-down or consensus-driven, we identify several findings. Since these two dimensions relate to **power-distance**, we often identify these three dimensions co-existing in messages. First, we see a clear hierarchy within the group. In several analysed messages, there is an individual called 'mango' who is reported as a mid-level manager (krebsonsecurity, 2022). Mango uses direct language with individuals in subordinate positions, indicating the strictness of the hierarchical structure. For example, *'..it was me who negotiated the amount of 4850 - well who asked you to give them such discounts ? they would have taken more from them'.* (2022-02-27T20:47:10.130398).

Similar behaviour is observed from other 'managers', e.g. 'tramp' who is direct and judgemental to subordinates. For example, a message on a numerical decision states:

*'you messed up there a couple of times so I decided 0.5'.* (2022-02-27T20:48:18.300051).

The **evaluating** dimension indicates direct feedback throughout the communications. Indicatively, around the timestamp '2022-02-22T17:29:39.918405' we observe several messages from 'pumba' to 'skippy' such as: *'they bit\*\*\*s messed up the last two ways'*, *'because that's how you fu\*\*\* find'*, and *'what the fu\*\* are you saying?'*. The Head (or CEO) of the group, 'stern', also provides direct negative feedback in the form of a warning: *'if you fu\*\* up one more time, I'll remove you from the toad forever'*[2].

In direct relationship with the hierarchical structure and **power-distance**, we observe various instances of direct and explicit feedback:

---

[2] Toad is a Database Developer and Administration Software Tool (https://www.quest.com/toad/).

*'..and you can tell derek that I don't give a f\*\*\* about your sentiments, I don't need it for you all the f\*\*\* time to excuse myself and decide anything'*. (2022-01-27T18:24:09.359805 ).

*'5% for negotiations, 20% for expenses and organization - for what? Don't you think it's funny?'*. (2022-01-27T18:24:09.357922).

The same pattern of **power-distance**, but with a reverse tone is confirmed by the communication of subordinates to the managers. The language used is more polite and indirect, as in the following message:

*'This is the first time I've had a forfeit in this job, so I wasn't expecting it, but I'll be more patient'*. (2022-02-15T18:00:43.559414).

In the same fashion, there are requests for members to be allowed a day off (2022-02-23T13:03:24.115653), and probationary periods with a reduced monthly salary (2020-06-23T11:22:50.816144).

Moreover, **hierarchy** and **power-distance** allow for the recognition of value of individuals within projects. Indicatively, the manager 'mango' expresses this to a group member, along with direct orders: *'show up already solve all these problems with me, there's no way without you now...'*. (2022-02-02T05:22:46.069845).

It is noteworthy that **leading** is not necessarily hierarchical, as we observe an egalitarian style of communication, at least within members of the leadership team. Although we connote be sure about the exact leadership structure, this finding indicates that power is not necessarily in the hands of a single person (namely, 'stern'), and that there is a group of individuals who govern the group. This finding is in line with the observed/purported structure of the group after its dissemination, i.e. that the Conti group became more of a conglomerate of cybercriminal groups, providing evidence for both a **leading** style which is egalitarian and a consensual **decision-making** approach within the senior members. Such distributed and shared leadership practices are common within online groups and communities (see, e.g., Chamakiotis and Panteli 2017; and Johnson et al. 2015).

Then, on the Hofstede **individualism-collectivism** and **masculinity-femininity** continuums, we have indications for close collaboration and sharing of information. The individual called 'arget' informs 'stern' about 'professor' who is purportedly a senior member, that *'prof fu\*\*s up every day and rolls somewhere'* (2020-10-18T05:59:26.382138). This indicates reduced information asymmetries within the group, as issues are escalated to the Head ('stern'), directly, in several instances. It can be argued that this indicates close collaboration at senior levels, thus, an aspect of **collectivism**, or even **femininity** within the group. However, masculinity and femininity contain various sub-variables, thus, this dimension cannot be clearly identified. But there are some indications, for example, the task-related instructions – often orders – and the direct feedback indicate a goal-driven environment (thus, high levels of **masculinity**) as well as a **task-based trust** mechanism within the group.

On the other hand, we see **collaboration** within the group, and, namely, sharing knowledge and expertise. In particular, members ask each other for help on various tasks. Some members communicate with their peers and discuss various topics, like vacation and food, indicating that co-worker and/or personal relationships are formed within the group:

*'What could be better than a breakfast of cold-cooked mackerel'.* (2022-01-02T04:11:56.322336).

The **communication** context is found to be low. That is, both requests and feedback are relatively laconic and to the point. Due to the identified high levels of power-distance, communication from subordinates to line managers is more contextual and more polite, however.

**Trust** is observed as a topic, as already mentioned. In some cases, managers provide direct feedback and indicate that the subordinate member has failed in previous assignments, and thus, they take some appropriate action. Therefore, trust-building is indicated as being task-based: *'you messed up there a*

*couple of times so I decided 0.5'*. (2022-02-27 20:48:18). Furthermore, trust associated with expertise and skills is observed. The finding highlights the perceived importance of key-roles within the group, with the angle of skills additionally indicating task-based trust relationships: *'..who can be trusted to manage coders for example'*. (2022-01-26T08:19:28.122588).

With regards to **uncertainty avoidance** and **long-term orientation**, there are some indications in the dataset, along with the protection of members. The most indicative example is the communications of 'mango' and 'kagas' with 'stern' on a case in the US; namely, members of the group were involved in an investigation and needed legal representation:

*'Alka is being transported from florida to ojai, she has a state attorney as she has no money for her own I understand. We can get the documents if our lawyer will make a defense agreement with her and will represent her. In order for him to start acting, we need to charge him 10k. And we need to figure out how to get it to the US safely... I don't know how and what to do next, everything is hanging there. Waiting for an answer.'* (2021-05-13 15:48:59). And subsequently:

*'Our old case was reopened, the investigator said why it was reopened, the Americans officially asked for information on Russian hackers, not only us, but in general who was caught around the country. In fact, they are interested in the trickbot, and some other viruses. The next Tuesday, the investigator summoned us for a conversation… Would it be possible to get our salaries? What if I could extend my lawyer and take a vacation till the end of October?'* (2021-10-06 11:50:36).

High UAI indicates a need for clear steps and instructions to be followed. On the other hand, long-term orientation indicates a pragmatic, holistic approach to handle difficult situations.

We also identify instances of **principles-first** persuading style. Statements such as '*that's not how we work*' (2021-03-08T22:24:34.789701) and *'I even gave him the salary no matter how much he fu\*\*ed me,*

*so he wouldn't pull his drunken bullshit on you..'* (2021-03-31T20:58:19.599765) indicates that deductive arguments are utilised in the communications.

**Themes**

The main finding identified from the analysis is power distance. Based on the individual identified dimensions, such as power distance, we explore combinations of dimensions into further constructs. In particular, we identify the theme of **connectedness** and a sense of **community**. In more detail, it is the dimensions of collectivism and long-term orientation which indicate this theme. The group is found to operate as a collective regardless of their different backgrounds and roles.

The finding of **power distance** and **connectedness** (collectivism) is reported in the literature as correlated (Hofstede et al., 2010, pp. 102-105). Countries with high scores of individualism are found to have low PDI scores, and vice versa, collectivist countries have higher PDI scores (negative correlation of individualism and PDI are found to span from -0.55 to -0.68; Hofstede et al., 2010; Oyserman et al., 2002).

Moreover, long-term orientation - as a component of connectedness - in combination with high PDI, indicate business characteristics such as, on the one hand fast adaptability to challenges (due to high PDI), and, on the other hand, patience, persistence, and investment in infrastructure, research, and generally, resources with slow return on investment (due to long-term orientation).

**Ideology** is not found to be present in the messages. In particular, ideology can be manifested via the co-existence of collectivism, high PDI, long-term orientation, femininity, restraint, and high UAI (see Table 1). The last three dimensions are not clearly identified. Thus, we cannot infer ideology as a dominant characteristic within the group.

**Risk-seeking behaviour** is also not explicitly identified, since masculinity is not clearly observed, nor is a low UAI. Equally, nor is risk aversion identified clearly, since it would require high scores in UAI, and a strong need for predictability and stability.

**Ethical attributes** are not identified in the analysis. Although collectivism is found, along with long-term orientation, the component of low PDI is not identified. Furthermore, neither consensual decisions, nor relationship-based trust is identified; the lack of evidence expands to egalitarian leadership and indirect feedback, implying that there is no identified evidence for this construct.

6. **Discussion, Limitations and Implications**

One of the realisations of this study is that a cybercriminal group is not a monolithic entity which makes rational decisions to maximise their profits. Instead it is a living organisation with several actors with their own motivations and approaches. However, a sense of community, shared identity and common goals appear to dominate over individualistic perspectives. The business-like environment and the everyday issues 'employees' face, portray a different image from the monolithic entity depicted in the news and online (indicatively, Meegan-Vickers, J., 2023).

The *task-based trust* mechanism identified might not be expected. The reason is that the group was international and the several members working in different cities and countries had to deliver their work remotely. Thus, the working conditions almost necessarily establish relationships – and trust – which are based on performance and accomplishments. The dimension, however, is noteworthy, since Russia is found to have a *relationship-based trust* culture, of similar levels with Mediterranean nations, but in contrast to, e.g., northern Europe (Meyer, 2015); this is potentially an additional argument for the international background of the group. Further, Russian culture has been reported as significantly different from Western countries (Elenkov, 1997; Beamish, 1992; Bollinger, 1994; Puffer, 1994).

On the other hand, a combination of sources along with internal communications indicate that the Russian element was dominant within Conti at the time of the messages leak. So, there is a discrepancy with regards to findings on trust, possibly caused by a business ethos and practices.

Instances of *deductive* arguments being used by different individuals are in accordance with the dominant Russian element, and thus, the culture, of the group.

With regards to the identified pair of high *PDI* and *collectivism* (as a component of connectedness), our findings are in line with published studies. Namely, Russia is found to have a remarkably high PDI score, and a collectivism index higher than the average country, but not exceptionally high. We identify connectedness in the analysis, and protective actions towards members, and, at the same time, there are instances of threats for firing members based on poor performance. We note that at the extreme end of collectivism individuals are not fired from businesses. The findings can be considered in line with the scores of Russia, i.e. moderately high collectivism and very high PDI.

An interesting association of collectivism is the fact that group decisions in strategic games are found to be more 'rational', i.e. with a traditional notion of rationality as expected value maximisation, in comparison to individual decision (Bornstein et al., 2003, p. 604). In combination with long-term orientation and risk seeking attitudes, this finding is in line with the 'business orientation' of the group.

We should note that *leadership* and *management* are often reported as 'cultures' of their own, implying that these cultures span across national cultures. It could be the case that ransomware groups, since they operate as businesses, have common characteristics which destroy the correlations with national cultural dimensions. Although such an argument might seem valid, existing research indicates that even seemingly unrelated aspects of social life are in fact related, i.e. linking national and business (organisational) cultures (Harris, 1981, p.8).

The business-related identified characteristics of connectedness and high PDI, are in line with Conti's evolution and actions, namely, calculated actions, investment in infrastructure and model, i.e. in terms of affiliations and business model, but also a relatively fast adaptability to changing circumstances, with the main example, the transformation of the group in 2022.

The lack of the ideological components might be in line with the business-like operations of the group, that is, a focus on objectives and profit. The Conti attacks on Costa Rica indicate an ideological disposition; however, their overall operations indicate the exact opposite. Moreover, assuming that the group's rebranding resulted from their support to the Russian government, the group's follow-up 'corrective' actions indicate a regret towards this announcement.

With regards to risk-taking, the group's modus operandi with multi-layered affiliate members, along with their success, indicate a long-term pragmatic orientation, and an overall avoidance of risk-taking which is in line with the finding that risk-seeking behaviour is not detected in the datasets.

On the importance and implications of the findings, some of the identified dimensions can indicate the characteristics of the dominant underlying national culture. Our findings confirm previous quantitative measurements where differences in power distance, collectivism, uncertainty avoidance, and machiavellianism have been reported to be statistically significant between American and Russian participants, and in this order of significance. Russian subjects are found to score significantly higher in all the above four dimensions (Elenkov, 1997). Thus, the Conti case study indicates that this kind of analysis can contribute to a broader framework of CTI to produce additional useful intelligence.

Evidence indicates a leadership structure shared amongst at least a few individuals. This is in line with existing literature on leadership in online communities and groups (e.g. Chamakiotis and Panteli, 2017). Nevertheless, there appears to be a strong presence of at least one leader who becomes central to community interactions and is able to exert influence on others as well as drive group activities.

Moreover, the reported transformation of Conti into a new entity indicates a similar hierarchical structure with several affiliate entities.

**Limitations and Implications for further research**

A first potential limitation is that the original messages were in Russian, whereas the analysis was conducted in a translated version in English. Although there is slang used in the messages, as one would expect, it does not alter the meaning of the communications, so in this case, language is a minor issue. Before conducting the analysis, we tested various sample messages with both online services (e.g., Google translate) and with Russian-speaking individuals, and we concluded that the translation of the provided dataset (TheParmak, 2023) is acceptable in terms of accuracy. We found that this dataset constitutes a more accurate translation in comparison to our own based on python, thus, we proceeded with this dataset.

The scores associated with the various cultural variables of Hofstede indicate relative positions, namely, they indicate a ranking order of, say, two countries with regards to power distance. The same holds for the Meyer variables. In that sense, our analysis could be utilised in comparing and ordering ransomware groups or even different types of cybercrime groups. This would allow for an understanding of the relative cultural positioning of the groups and might allow for identifying similar characteristics and patterns between types of groups and cultural dimensions. For our purposes here, we provide a case study of the analysis on a single group, to showcase how the analysis is conducted, and what potential findings can reveal.

Within the dataset, there were also various messages which were encrypted; consequently, we did not have access to these for the analysis, thus, limiting the depth and breadth of insights in specific topics.

The roles and relative position in the hierarchy is identified for several individuals within the group. However, a more detailed knowledge of the structure would be helpful, since it would allow to control for position-related behaviours in more detail, e.g., the way that a subordinate member communicates with someone higher up in the hierarchy. That is, such information would allow for a 'noise reduction' in the data, especially with regards to personal dispositions and communication styles. The social context, however, is clearly observed, as in the case of observing the dimensions of deciding, power-distance, leadership, and the social norms within the group.

## 7. Conclusion

The main aim of the paper has been to showcase the usefulness of cultural research in cybercrime investigations. By analysing the Conti ransomware group leaked datasets across 14 cultural variables, we identify dominant cultural dimensions and constructs within the group, namely, high power-distance and connectedness. We portray how such analysis can provide insights on cybercriminal operations and suggest that our approach can be embedded in cybercriminal profiling frameworks, by linking our findings with information on Conti, and by connecting them with reported correlations of cultural dimensions in the literature. In combination with extended sources of CTI, our proposed approach can contribute to cyber-attack attribution providing a useful apparatus for law enforcement and nation states.